# Mesoscopic Models of Plants Composed of Metallic Nanowires

Galina K. Strukova, Gennady V. Strukov, Evgeniya Yu. Postnova, Alexander Yu. Rusanov, Ivan S. Veshchunov

*Institute of Solid State Physics, Russian Academy of Sciences, 142432 Chernogolovka, Russia*

**Abstract**

Various metallic structures of complex shape resembling living plant organisms (biomimetics) are produced as a result of self-assembly of nanowires growing on porous membranes in the course of pulse current electrodeposition. These structures occur if the electroplating is continued after the nanowires appear on the membrane surface. By varying the membrane geometry, pulse current electroplating parameters, and alternating electrodeposition from two baths composed of a variety of electrolytes, diverse models were fabricated, including a hollow container with a wall thickness of 10 nm – 20 nm. This biomimetic method suggests an analogy between the shape-forming processes of plants and their metallic models. Nanostructured mesostructures of various metals (Ag, Pd, Ni), alloys (PdNi, PbIn) and hybrid structures (PdNi/Pb, PdNi/PbIn) were obtained. They can be of interest for fundamental research (self-assembly, morphogenesis) as well as for applications in nanotechnology (catalysis, nanoplasmonics, medicine, superhydrophobic surfaces).

## 1 Introduction

Recently, methods of producing bio-inspired models have become an increasingly urgent research issue in nanoscience and nanotechnology[1]. Researchers' interest has been attracted by nanoflowers — anisotropic semiconductor and metal nanostructures. For instance, hyperphotoluminescence on zinc oxide-based nanoflowers[2,3], the effect of giant Raman scattering on gold[4,5], and catalytic activity on platinum nanoflowers[6,7] have been reported. However, the known isolated cases of synthesis of metallic models of natural objects, including nanoflowers, provide no information on general methods that enable synthesis and shape control of diverse structures. Is there something common in these separate syntheses of metal "nanoflowers"? What causes the plants' shapes? Morphogenesis mechanisms of biological objects have always attracted scientists' attention[8–11]. We investigated biomimetics through experiments on nanowire growth in a porous membrane by electrodeposition, obtaining different metal "plants". Electrodeposition in porous membranes is one of the solid methods for growing metal nanowires[12,13]. Usually, the potentiostatic process is terminated right after the appearance of nanowires on the membrane surface, with subsequent dissolving of the host material in order to isolate particular nanowires. In our case, pulse electroplating was used and the process of growing nanowires was continued after the nanowires appeared on the membrane surface. This resulted in various nanostructured models resembling living plant organisms.

The aim of the present work is to demonstrate the possibility of the controlled growth of plant-like metal nanostructured models and the variety of their shapes. The details of techniques for growing the structures and the general physical properties of the obtained models will be discussed elsewhere.

## 2 Experiments

We prepared metallic mesostructures on porous membranes via pulsed current electroplating using a simple two-electrode scheme. Aluminum oxide membranes 15 mm wide and 60 μm thick with through-pores of typically 200 nm, 100 nm, and 50 nm and polymer membranes 23 mm wide and 53 μm thick with

**Corresponding author:** Galina K. Strukova
**E-mail:** strukova@issp.ac.ru

through-pores of 100 nm and 50 nm were used as templates. The membranes were covered with 50 nm –100 nm copper layer that served as a cathode in the process of electrolysis. To grow nanowires only through the pores, the cathode backing was covered with polytetrafluoroethylene to prevent any undesired electrical contact with the electrolyte. Platinum foil, placed several millimeters in front of the membrane, was used as an anode. The cathode and anode were electrically linked to the pulsed current generator and placed in a bath with the electrolyte to undergo a cycle of current pulses.

In order to grow volume metal structures with the porous membranes as templates, aqueous solutions of electrolytes were prepared. The silver-plating solution contained (g/L): $AgNO_3$–40; sulfosalicylic acid $C_7H_6O_6S \cdot 2H_2O$–105; $(NH_4)_2CO_3$–25; $(NH_4)_2SO_4$–70. The electrolyte for lead electroplating contained (g/L): lead acetate $Pb(CH_3COO)_2 \cdot 3H_2O$–50; disodium EDTA –51. The nickel-electroplating solution contained (g/L): $NiSO_4 \cdot 7H_2O$–220, $NiCl_2 \cdot 6H_2O$–60, $H_3BO_3$–35. The electrolyte for Pd electroplating contained (g/L): $PdCl_2$–17; $NH_4Cl$–55, sulfamic acid $HSO_3NH_2$–75; $NaNO_2$–60; aqueous solution of $NH_3$ – up to pH = 8–9. The electrolyte for PdNi alloy electroplating contained (g/L): $PdCl_2$–6; $NiCl_2 \cdot 6H_2O$–130; $NH_4Cl$–75; ammonium sulfamate $NH_4SO_3NH_2$–100; $NaNO_2$–30; aqueous solution $NH_3$ – up to pH = 8. The electrolyte for PbIn alloy electroplating contained (g/L): $PbCl_2$–33; $InCl_3$–77; disodium EDTA–75; aqueous solution $NH_3$ – up to pH=7. For PbBi alloy electroplating, we used aqueous electrolyte based on a trilonate complex of lead and bismuth[14]. The electroplating of Ni and PdNi alloy was carried out at temperatures of 35 ˚C – 50 ˚C, while all other metals and alloys were electroplated at 20 ˚C – 25 ˚C.

It is necessary to point out that the electroplating of Ag, PdNi, and PbIn alloys was done with a particular electrolyte, whereas, in the cases of PdNi/PbIn and PdNi/Pb electroplating, two baths with corresponding electrolytes were used alternately[15]. Table 1 represents the main parameters of the pulsed current electroplating procedure used for growing the particular structures.

Scanning Electron Microscope (SEM) investigations of the obtained samples were carried out on a SUPRA 50VP scanning electron microscope.

## 3 Results and discussion

Fig. 1 shows a cross-section of the membrane with formed nanowires inside and the volume structures that appeared on the membrane surface. In Fig. 1, we can see a gibbous structure "cauliflower" made of PdNi alloy nanoclusters and a complex structure "fern" obtained by alternating electroplating of PbIn and PdNi alloys.

### 3.1 Mesostructures of Ag, Pd, Ni

Different Ag-structures are shown in Fig. 2. Structures in Figs. 2a–2c were formed on the Anodic Aluminum Oxide (AAO) membrane with a disordered arrangement of pores. The distinctive structure of the "branches and berries" (Fig. 2a) contains long elements ("branches") with spherical or elongated granules of silver sized 20 nm to 40 nm ("berries"). The structure "branch with leaves" (Fig. 2c) is decorated by swept flat "leaves" (Fig. 2b). The convex-concave structures that mimic the form of cabbage leaves were grown on the polymer membrane via slow growth from the diluted 1:1.5 electrolyte (Fig. 2d). A complex "inflorescence" from palladium was obtained on the polymer membrane (Fig. 3).

**Table 1** The main parameters of the pulsed current electroplating. $I$ is the amplitude of passing current pulses, $n$ is the number of pulses passed, $v$ is the pulsed current frequency, d.c.= 100 Ton /(Ton + Toff)%, (where Ton–pulse duration, Toff – pause between pulses) is the pulse duty-cycle

| Material | Membrane used | Pore diameter (nm) | $I$ (mA) | $n$ | $v$ (Hz) | d.c. (%) | Shown in (Fig.) |
|---|---|---|---|---|---|---|---|
| PdNi | Oxide membrane | 100 | 60 | 4000 | 30 | 90,9 | 1a |
| PbIn/PdNi* | Oxide membrane | 200 | 30/30 | 500/1000 | 9,7/43,5 | 97/87 | 1b,7 |
| Pb/PdNi* | Oxide membrane | 200 | 30/30 | 500/1000 | 9,7/43,5 | 97/87 | 8 |
| PdNi | Polymer membrane | 50 | 100 | 30000 | 30 | 91 | 6 |
| Ag | Oxide membrane | 100 | 50 | 40000 | 56 | 83 | 2a |
| Ag | Oxide membrane | 100 | 20 | 40000 | 17,9 | 94,6 | 2b, 2c |
| Ag | Polymer membrane | 50–100 | 50 | 40000 | 77 | 77 | 2d |
| Pd | Polymer membrane | 50–100 | 200 | 35000 | 71 | 79 | 3 |
| Ni | Polymer membrane | 50–100 | 200 | 35000 | 77 | 77 | 4 |

*The process was carried out for 20 cycles each of which contained 500 pulses for PbIn (Pb) and 1000 pulse for PdNi.

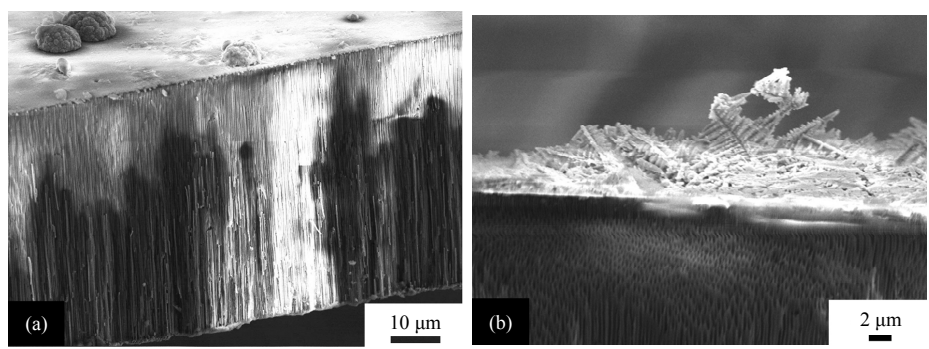

**Fig. 1** Cross section of the membrane with nanowires inside and structures on the surface: (a) "Cauliflower" of PdNi alloy; (b) "fern" of PbIn and PdNi alloys.

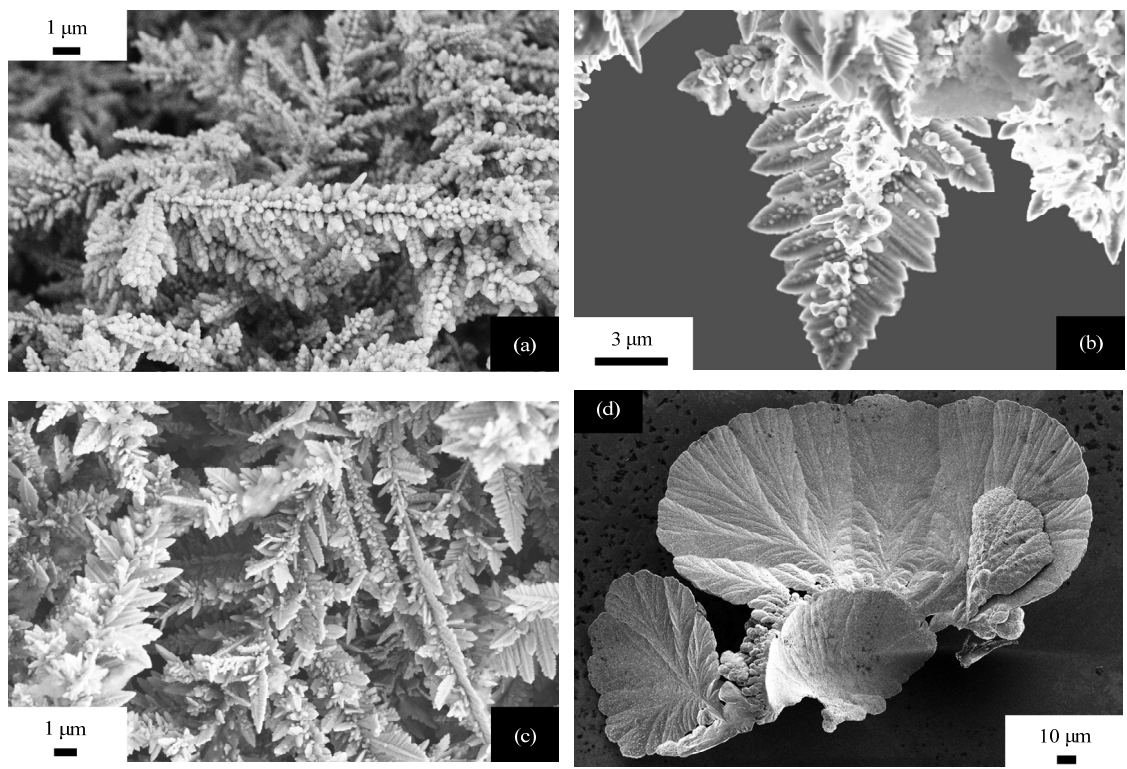

**Fig. 2** Ag structures: (a) "Branch with berries"; (b) distinctive "leaf"; (c) "branches with leaves"; (d) "cabbage leaves".

When using nickel-plating electrolyte under equal experimental conditions, different sphere-like structures, single and consisting of several spherical elements (Fig. 4), were obtained as a result of fractal branching. Spheric structures resembling mushrooms grow from a "root" (Figs. 4a and 4b). The inside of the spheroids displays both patterns formed by self-organizing arrangment of nanowire bunches ("meridians" - Figs. 4c and 4d) and lines corresponding to the changes of current pulse series ("parallels" - Figs. 4a–4c). The outward surface of the spherical mushroom is formed by nickel submicron crystallites. Their morphology corresponds to the typical morphology of nickel axis-oriented (211) crystallites that are obtained from this electrolyte with pulsed current electroplating on the substrate.

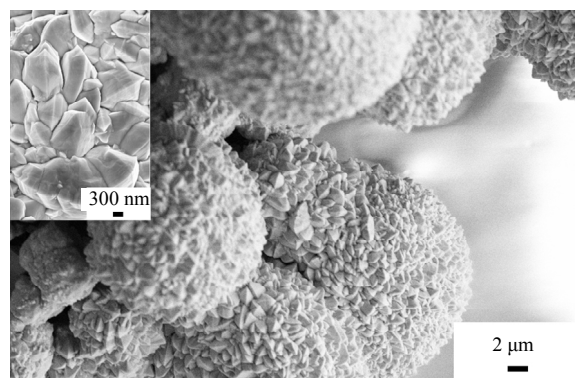

**Fig. 3** Mesostructure of palladium.

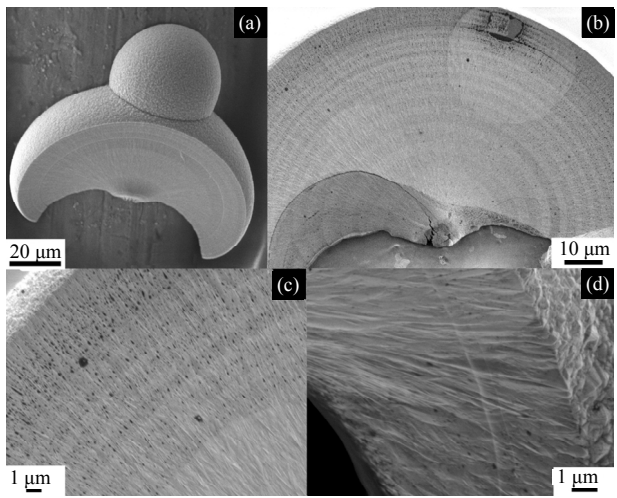

**Fig. 4** "Mushrooms" of nickel.

### 3.2 Mesostructures of PdNi alloy

Electrodeposition of PdNi alloy on porous aluminum oxide membranes with disordered pore arrangement gave rise to diverse "vegetable" structures ("cauliflower", "broccoli", "squash"), ensembles of densely packed nanowires in the form of agaric or fungus on trees, as well as elegant convex-concave shell-, lotus leaf-, and mushroom-like structures. Some of them are presented in Fig. 5. It was observed that, quite often during the electroplating procedure, small bunches of nanowires appearing locally on the membrane surface simultaneously evolved into a larger community, continuing their growth together and forming nanosized and submicron "algae" with flower-like buds on their tips.

Depending on membrane pore size and geometry as well as on electroplating regime, particular groups of nanowires morph at different stages of their growth into "bouquets" (Figs. 5f and 5g) and "patty-pan squashes" (Fig. 5h). The size of such structures may reach dozens of microns.

We reproduced the convex-concave structures using specially prepared polymer membranes in combination with the found electrodeposition modes. Fig. 6 shows a number of these models ("lotus leaves") prepared in the found conditions. These experiments demonstrate the possibility of shape regulation for models.

Energy-dispersive X-ray (EDX) analysis gave $Pd_{72}Ni_{28}$ composition for those structures.

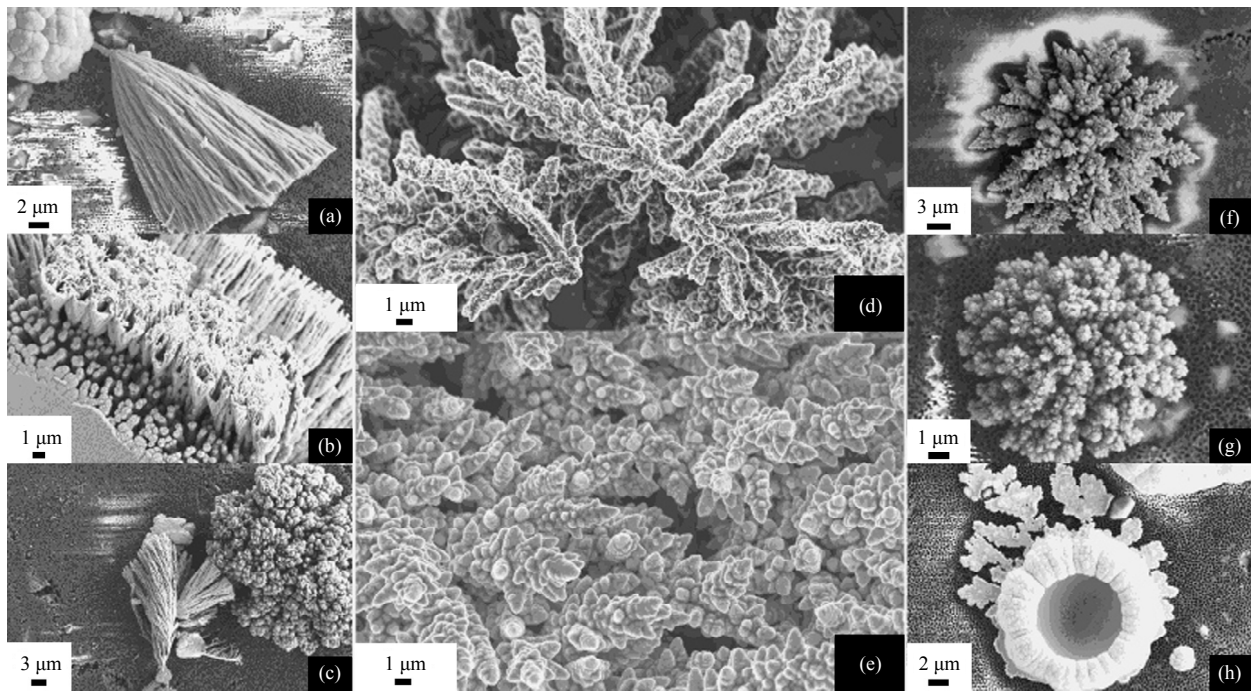

**Fig. 5** Volume PdNi alloy structures on the membrane surface: (a) "Cauliflower" and "algae"; (b) "vegetable" structures at different stages of growth; (c) "brooms" and "broccoli"; (d) "cactus"; (e) "coniferous branch"; (f) and (g) " bouquets" and (h) "patty-pan squash".

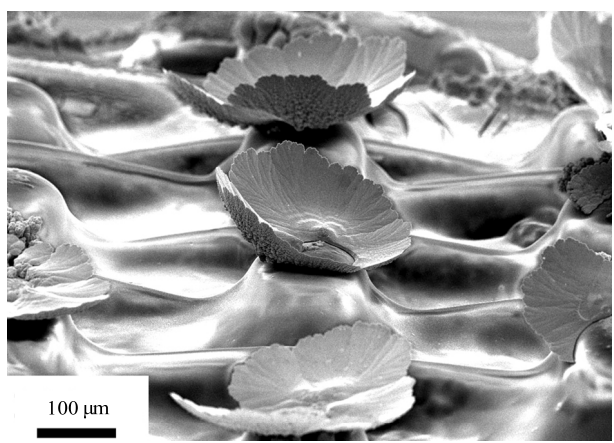

**Fig. 6** PdNi alloy. Rows of "lotus leaves" grown from specifically prepared pores.

### 3.3 Hybrid structures PdNi/PbIn and PdNi/Pb

A special technique of alternating electroplating of PbIn and PdNi alloys from two baths[15] allowed us to obtain a volume hybrid structure "fern". Fig. 7 demonstrates this model and its fragments. One can see replication of the complex order in these structures. Arrow-like leaves formed by four orthogonal blades are situated along the stem; the blades consist of alternating submicron columns formed by 100 nm – 200 nm spherical grains.

A similar scenario, shown in Fig. 8, occurred when electroplating Pb and PdNi alloys alternately. Fig. 8a shows long straight "palm branches" with straight "leaves" (also see Fig. 8b), each tipped with a wonderful sedum-like bud. Also formed was a hollow tube with a wall thickness of 10 nm – 20 nm (Fig. 8c). EDX analysis indicates the presence of all metals in the fragments of the grown hybrid structures (Pb, In, Pd, Ni in "fern", and Pb, Pd, Ni in "sedum"), yet in order to find the local distribution of PbIn (Pb) and PdNi alloys a technique with nanometer resolution is required.

### 3.4 A possible mechanism of "vegetable" metallic model formation

The method of producing nanowires by DC electrodeposition of metal on a porous membrane at controlled potential has long been known and extensively used. As a rule, the process of nanowire growth is finished immediately upon nanowire rise to the membrane surface, following which the membrane is dissolved and the nanowires are isolated for targeted use. If the process is continued after nanowire rise, the membrane surface is healed over and used as a contact pad for measuring the wire resistivity. In this case there are no mesostructures.

There are two scenarios possible for growing metal nanowires on porous membranes by means of pulse current electrodeposition. The first one is when nanowires, after they have appeared on the membrane surface, continue to grow separately. The second one, considered in the present paper, is when nanowires form nanostructured "vegetable" metallic meso-samples on the membrane surface. This occurs in the case of self-assembly of nanowires that appear on the membrane surface simultaneously with relatively small distances between the nanowires. Both methods were implemented by us.

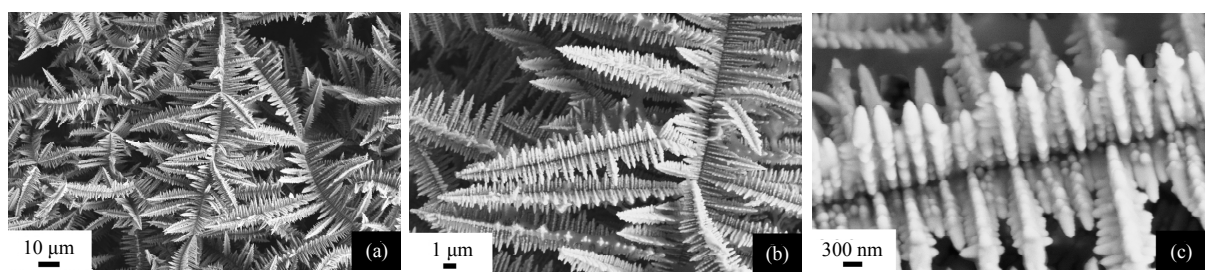

**Fig. 7** "Fern" of PbIn and PdNi alloys. (a) General view; b) "fern leaves"; (c) "fern" stem.

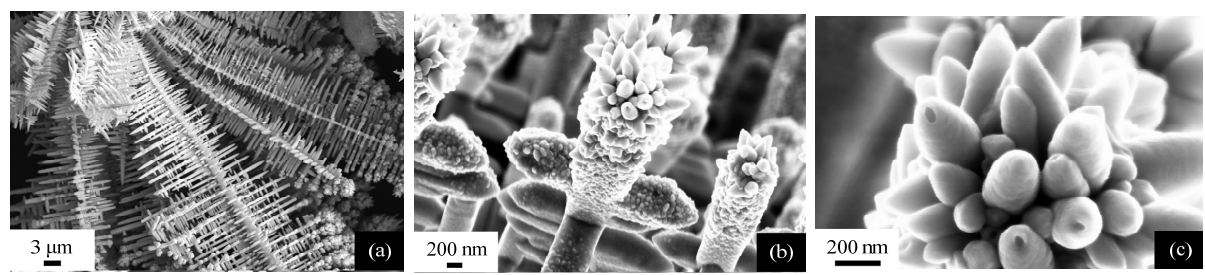

**Fig. 8** "Flower garland" of Pb and PdNi alloys. (a) General view; (b) "inflorescence"; (c) "buds".

The results presented show that pulse current electrodeposition on porous membranes ensures controlled growth of nanostructured metallic models mimicking natural objects, plants and fungi, i.e., it is a biomimetic method of their synthesis. Pulse current is responsible for discrete addition of material, here, in cluster form. A distinctive feature of pulse current electrodeposition is the feasibility of realizing the process at a higher current density than in the case of DC electrodeposition, thus ensuring nanocrystalline metal deposition. It is also known that the use of a programmed pulse current mode considerably enhances control of the structure of the deposited coatings, as compared to the galvanostatic and potentiostatic conditions which are related to the differences in the electrocrystallization kinetics[16].

The template is another important component of the method in question. Electrodeposition on smooth surfaces at different pulse current modes does not lead to the formation of "vegetable" mesostructures. Fig. 9a shows a PdNi coating obtained on a copper foil from the same electrolyte and at the same pulse current mode as the "vegetable" structures. The coating is solid, smooth, and nanocrystalline, with grain size 30 nm – 70 nm. The diffraction pattern of the coating reveals that it is a solid solution and the grain size, calculated from the peak broadening by the Selyakov-Sherer formula, is 20 nm – 35 nm. Surfaces with deposited metallic patterns or etch pit edges can serve as templates. For example, when a foil was pre-etched, then, at the same pulse current mode, we observed the growth of nanowire ensembles (Fig. 9b). Membrane nanowires serve as the first template during the growth of "vegetable" mesostructures. Self-assembly of metallic nanoclusters deposited on the template results in the formation of elementary "bricks" acting as templates for deposition of following elements. Thus, the combination of pulse current and a template ensures the formation of "vegetable" structures, i.e., biomimetic model growth, and is a tool for hierarchical structure formation at the nano-, micro- and meso-levels. This is confirmed in the literature by some examples of growing "nanoflowers".

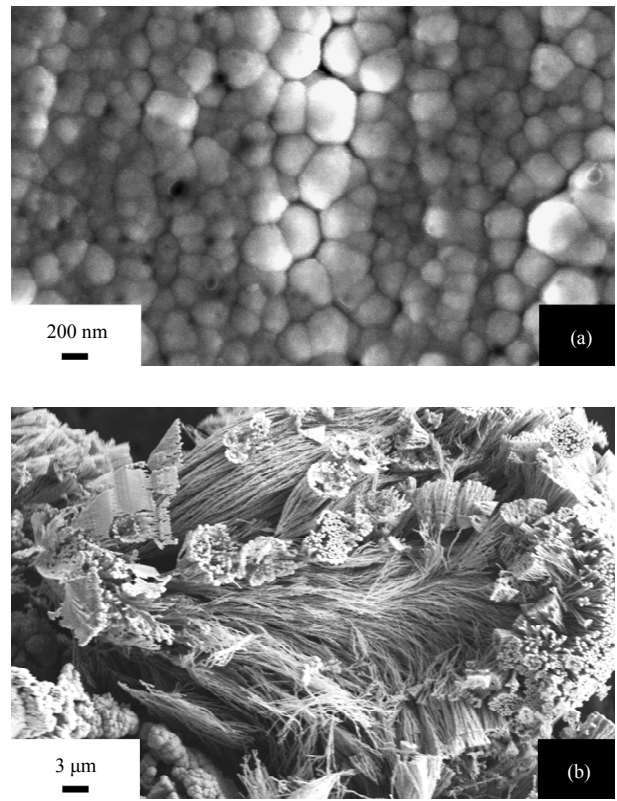

**Fig. 9** PdNi alloy coatings obtained from the same electrolyte and at the same pulse current mode: (a) On the smooth Cu-foil; (b) on the etched Cu-foil.

In the literature, the typical work on the synthesis of "nanoflowers" (by chemical deposition and electrodeposition) reveals one common feature: the material is deposited on an as-prepared porous membrane[3], or else deposition is preceded by formation of a porous polymer template (matrix)[17], a honeycomb metal frame[5], or a zinc oxide sublayer with specific morphology[2]. These facts suggest that the known examples of "nanoflowers" represent a special case of the general biomimetic method of creating models during irregular (pulse) growth of material on porous membranes or templates. For instance, in Ref. [17] the authors grew a silver "leaf" on a polymer template by chemical reduction from solution, a leaf which is identical in form to our model (Fig. 2b) grown on an oxide-aluminum membrane from another electrolyte by pulse current electrodeposition. It is clear that, in the above cases, the irregular pulse deposition mode was realized spontaneously owing to the concentration inhomogeneities developed in the reaction system. Other separate cases of grown "nanoflowers" displayed no common biomimetic method that enables shape modeling.

### 3.5 The analogy with the process of plant shape formation

The shape resemblance of the synthesized metallic mesostructures to biological objects is obvious. As shown above, the synthesized metallic models also exhibit a hierarchical structure inherent in natural objects. We believe that the shape resemblance of our grown metallic mesostructures to biological objects is not accidental. It is likely to be explained by the coincidence of some essential mechanisms of the growth processes and shape formation. It is known that the spatial orientation of growth, that is, the polarization of biological tissues, is caused by multiple factors: osmotic gradient pressure, pH, concentration of oxygen and carbon dioxide, hormonal, electrical and trophic contact with neighboring cells, force of gravity. With respect to the analogy between biological objects and the metallic models we investigated, we can identify several important general factors of both the growth processes and the shape formation of plants and their metallic models.

(1) Growth occurs on membranes or templates.

(2) Pulsed or irregular growth.

(3) Growth is achieved by addition of material at certain growing points, the so-called "growing tops" for plants and the "nuclei" in the case of electrocrystallization. The location of the growing points is shape-determining and leads to fractal branching. It should be noted that practically all the grown metallic mesostructures are fractals. They are classical Mandelbrot fractals, e.g., the "cauliflower" and "broccoli"[11]. "Fern" and "sedum-like buds" are also well-expressed fractals. This aspect of growing metallic models by pulse current electrodeposition on porous membranes is also similar to biological morphogenesis.

Considering that the formation processes of natural objects and their metallic models have similar factors, and that the diverse synthesized models have both external and structural resemblance to the biological prototypes, we can see the general instrument of morphogenesis. We suppose that pulsed growth on membranes is a tool of morphogenesis for most natural mushrooms and plants, and that it is accompanied by fractal branching and self-assembly of growing clusters and fibers. The species diversity in nature is likely to be due to the diversity of biological membranes that are known to be synthesized in the course of self-assembly involving genetic codes and various pulsed growth modes. The latter, in turn, may depend on numerous factors such as solar intensity, water amount, chemical growth-promoting agents, *etc.*

### 3.6 Possible practical applications

In the case of synthesized metallic mesostructures, their diversity is due to the combination of membranes with different pore geometries and different algorithms of nanocluster growth. Membrane geometry implies the shape of pores and their distribution on the membrane. Nanocluster growth is determined by current amplitude, pulse and pause duration, as well as by the wide range of diverse pulse-pause combinations. Clearly, our method enables the possible creation of a much wider variety of structures than presented here. The use of definite membrane geometry, along with a programmed mode of electrodeposition of different metals and alloys, allows for the control of the formation and reproduction of nanostructures and would be of interest for applications in nanotechnology.

Undoubtedly interesting to researchers are nanosize hybrid SFS-structures that open up prospects for superconducting digital electronics[18]. This calls for the investigation of the local distribution of superconducting (PbIn) and magnetic (PdNi) alloys and their boundaries in hybrid "fern" mesostructures. The creation of self-assembled SFS-structures has an attractive technological prospect.

Due to its technological simplicity, template growth of nanostructured metal and alloy coatings may compete with other methods of fabricating superhydrophobic surfaces for technical applications.

The nanostructure, the large surface area, and the conducting nanoarchitecture are the three major properties that make our metallic mesostructures most promising for applications in electrocatalysis, batteries and electrochemical capacitors[19–21]. The presented method is suitable for the fabrication of large surface area structures; some are shown in Fig. 10.

A forest of silver (Fig. 10a) could be used in medical filters. The tubular metallic elements and their ensembles could be used in nanotechnology as reactors and containers.

The method proposed enables the growth of nanostructured composites of metal-semiconductor (cadmium sulfide) mesostructures. This material holds

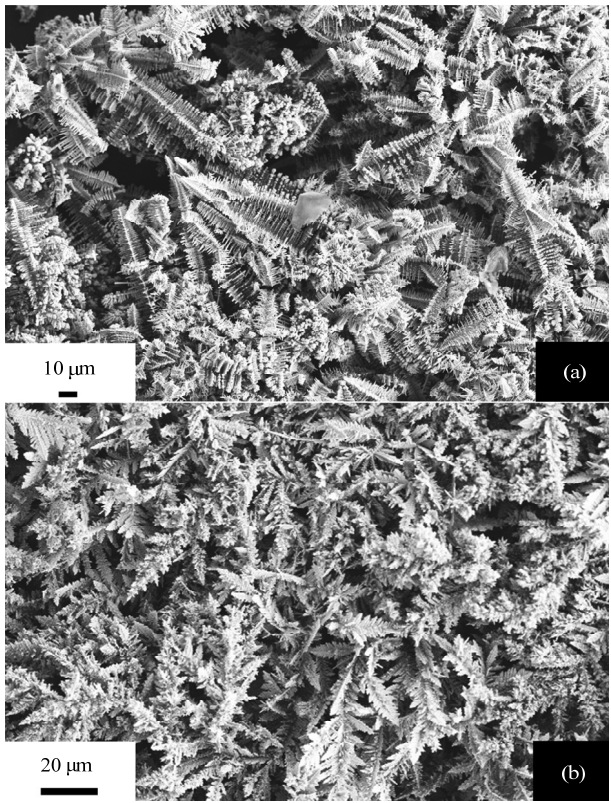

**Fig. 10** The metallic nanostructured mesostructures. (a) Ag; (b) PdNi/Pb.

promise for a practical use in solar energy production, e.g., via plasmon-enhanced catalysis[22].

PdNi-based nanostructured mesostructures could be used as efficient catalysts for hydrogenation of unsaturated hydrocarbons, for instance, in the hydrogenation of vegetable oils (margarine production).

## 4　Summary

Pulse current electrodeposition on a template is a biomimetic method for the fabrication of metallic models. By varying pore size and pore pattern in a membrane, and by varying the electrolytes and the pulsed current parameters, an impressive manifold of metallic nanostructured mesostructures resembling natural objects (mushrooms and plants) can be produced.

The forms of the metallic mesostructures can be regulated: the conditions for the synthesis of mushroom- and lotus-like convex-concave models via self-assembly of nanoclusters and nanowires were found. The growth factors and shape formation of the natural objects are analogous to those of the metallic models. These general factors, as well as the diversity of the synthesized models and their external and structural resemblance to the biological prototypes, suggest that pulsed growth on membranes is also a tool of formation of many plants and mushrooms.

The results obtained are certainly of interest for the development of nano-scale self-assembly processes as well as for the simulation of biological morphogenesis. Due to the technological simplicity, the grown nanostructured entwined metal materials offer considerable promise for nanoplasmonics, for the fabrication of efficient catalysts, for superhydrophobic surfaces for technical applications, for medical filters, and for sensors. The obtained structures can be grown using superconducting, ferromagnetic, and normal metals and, therefore, are potentially interesting for applications in building nanodevices. The manifold of structures that can be created by our method is significantly broader than discussed in the present article. Using membranes of certain geometry together with programmed regimes of electroplating of various metals and alloys, the creation and reproduction of nanostructures could be controlled, which could be of interest to researchers and engineers of nanodevices.

## Acknowledgment

The authors thank engineer Aegyle Shoo for his help in this work.